\documentclass[aps,pra,twocolumn,showpacs,floatfix]{revtex4}
\usepackage{epsfig}
\usepackage{graphicx}
\usepackage{dcolumn}
\usepackage{longtable}
\usepackage{amsthm,amsmath}

\begin{document}

\preprint{\today}

\title{Plasma-screening effects in the atrophysically relevant He-like and Li-like Mg and Fe ions}

\author{B. K. Sahoo \footnote{Email: bijaya@prl.res.in}}
\affiliation{Atomic, Molecular and Optical Physics Division, Physical Research Laboratory, Navrangpura, Ahmedabad - 380009, India}
\author{Madhulita Das \footnote{Email: physics.madhulia@gmail.com}}
\affiliation{Department of Physics and Astronomy, $^1$National Institute of Technology, Rourkela, Odisha, India}

\begin{abstract}
  The effect of plasma environment on the atomic energy levels of He-like and Li-like Mg and Fe ions have been studied using 
Debye model. The equation-of-motion coupled-cluster (EOMCC) and Fock-space coupled-cluster (FSCC) formalisms  in the relativistic
frame work have been adopted to describe the atomic states and the energy levels of the above plasma embedded ions. Salient features 
of these methods have been described 
to account the two electron screening effects through the Debye potentials. The two-body screening potential has been derived in 
the multipole expansion form to evaluate the reduced matrix elements in solving the equation of motion. Using this extended model, we
have also predicted 
that quasi-degeneracy among the energy states having same principal quantum number ($n$) but different angular momentum ($l$) is 
slacken, whereas fine structure splitting is unaffected with increasing plasma strength. These knowledge are useful in estimating
radiative opacity, photoionization cross sections, line intensities, etc of the aforementioned astrophysical plasmas.

\end{abstract}


\maketitle

\section{Introduction}
The emission spectra of helium-like (He-like) and lithium-like (Li-like) ions have been observed in recent years both in 
astrophysical and laboratory plasmas. The X-ray spectra from these ions are observed from a large variety of astrophysical 
sources \cite{mehdipour,nahar,schulz} and laser plasma interactions \cite{boehly,schaumann}. They are used to determine the
primordial abundances of elements that are of immense interest for testing the big bang cosmology \cite{silver}. Detection of 
these lines using astronomy telescopes reveal many X-ray sources. Also, the Earth's atmosphere absorbs most of the radiation 
at the X-ray wavelengths, so its behavior and dynamics can be investigated by knowing these lines accurately. Accurate
determination of these lines have become very demanding in recent years with the advent of increasingly powerful X-ray 
satellite telescopes. The two most powerful telescopes to date are the Chandra X-ray observatory, launched by NASA, in 1999
\cite{schwartz}, and the X-ray Multi-Mirror (XMM) Newton telescope, launched by the European Space Agency, 
recently \cite{lumb}. The ASTROSAT satellite, launched last year by Indian Space Research Organization (ISRO) is also aiming at 
exploring the possible X-ray sources in the space \cite{astrosat}. The international X-ray observatory is being planned to be 
launched in 2021 jointly by NASA, European Space Agency (ESA) and Japan Aerospace Exploration Agency (JAXA) which can cover a 
large effective area to probe for the X-ray sources \cite{nasa}. This is why X-ray astronomy remains to be an active research area 
today.  Due to high spectral resolution and sensitivity of the current generation X-ray satellites Chandra and XMM-Newton, it is 
possible to resolve the He-like ion lines distinctly and use them in the diagnostics for extra-solar objects. Indeed, the He-like 
ion line ratios are valuable tools in the analysis of high-resolution spectra of a variety of plasmas such as Collisional 
Ionization Equilibrium (CIE) plasmas or also called coronal plasmas \cite{mehdipour}. In such plasmas, ionization occurs due to 
electron-ion collision processes and the atomic levels are populated mainly by the electron impact. It is commonly assumed that 
CIE plasmas are optically thin to their own radiation, and there is no external radiation field that affects the ionization 
balance. However, in some cases, these assumptions are not fulfilled. In this case, recombination-dominated or Photo-Ionization 
Equilibrium (PIE) plasmas play the important role \cite{mehdipour}, where ionization takes place due to photons (ionizing 
radiation). As a result, the atomic levels are populated mainly by the radiative recombination processes directly or by 
cascading from the upper levels. These plasma are generally over ionized relative to the local electronic temperature and have a 
much smaller electronic temperature compared to CIE plasmas. That is why collisional excitations out of the ground state are 
inefficient and the excited levels are populated via the radiative recombination processes.  

Accurate knowledge of spectral lines of plasma embedded ions have become increasingly important today due to their observations 
using the high resolution detectors of the space-based X-ray observatories. Especially, the highly forbidden ``triplet'' 
inter-combination, and resonance lines of the He-like and Li-like ions have been used to measure temperature and density of ions
and electrons in the solar corona \cite{porquet, gabriel1,gabriel2}. Compared to other ionic iso-electronic sequences, He-like 
ions are abundant over the widest temperature range in collisional plasmas due to their closed-shell ground state 
\cite{mehdipour,boehly,porquet,moitra}. The most intense He-like lines correspond to transitions between the $n = 2$ shell and 
the $n = 1$ ground state shell \cite{moitra,bailey}, where $n$ is the principal quantum number. These He-like lines were first 
observed in laboratory for C, F, Mg, Al (see \cite{edlen}) and later in solar plasmas by the Orbiting Solar Observatory (OSO) 
\cite{sylwester} and rocket experiments \cite{fritz,acton,doschek1}. Gabriel and Jordan had argued for use of a suitable 
theoretical many-body method for the identification of the wavelength of the transition from the metastable ${^3}S_1$ level to 
the ground level \cite{gabriel1} and later Griem had demonstrated it using the quantum-relativistic calculation \cite{griem}. 
Since the pioneering work of Gabriel and Jordan \cite{gabriel1}, several works have been dedicated to the improvements of these
diagnostics based on spectral lines of the He-like ions and their extension to other types of plasmas (PIE and non-ionization 
equilibrium \cite{porquet1,bautista,porter,smith}. Gabriel and Jordan had also proposed that the relative intensities of these 
lines can be used for temperature and density diagnostics for solar plasma \cite{gabriel2}, which have been widely used for solar 
spectra \cite{acton,doschek1,doschek2,wolfson,doyle} and for X-ray spectra of tokamak plasmas \cite{keenan,kaellne}. 
Some astrophysical plasmas, such as supernova remnants, solar and stellar flares, colliding winds in star clusters and X-ray 
binaries, cluster of galaxies, intra-cluster medium in merging galaxy clusters, etc., depart from the ionization equilibrium 
when one or several physical conditions like temperature, electron or ion density and photo-ionization radiation field of the 
plasma suddenly change.

The absorption lines of Li-like ions occur close to the emission lines of the triplets of their corresponding He-like ions. The 
absorption lines of these ions can, therefore, affect the intensity of the triplet lines. In particular, those of the 
inter-combination lines as shown in the present study. Without taking into account in the analysis the absorption lines of 
the Li-like ions intrinsic to a medium emitting lines close to wavelengths of the triplets in the He-like ions, it can mislead
to wrong plasma diagnostics \cite{mehdipour}. Transition lines from Li-like Fe are quite useful for plasma diagnostics 
\cite{schulz}. Recent laboratory studies have shown that the intensity of the 3d-4f and 3d-5f lines in the Li-like ions increase
exponentially with Debye length in experiments conducted using very low power laser pulses \cite{boehly}.

In the present work, we would like to focus on investigating transition lines in the He-like and Li-like Mg and Fe ions in the plasma 
environment with ion and electron densities of the order of $10^{21}$ cm$^{-3}$ and for the temperature range 0 to 150 eV. Some of these transitions 
were experimentally verified in a recent work \cite{bailey}, but we shall provide these spectra for a large number of transitions which are 
mostly lying in the X-range regime. The paper is organized as follows: In Secs. \ref{sec2}, we introduce the screening models 
that are considered in the calculations for the description of the atomic spectra and Sec. \ref{sec3} describes employed methods for calculations briefly. 
In Sec. \ref{sec4}, we present the results, compare with the other studies and discuss them before summarizing the work in Sec. \ref{sec6}.
Unless stated otherwise we have used atomic units (a.u.) through out of this paper.

\section{Debye Model for Plasma Screening}\label{sec2}

The plasma environment mostly consists of ions and free electrons, which introduce screening effects in the Coulomb potentials of the embedded 
atomic systems. As a result, the atomic electrons are highly influenced by the external electromagnetic fields compelling the atomic long range 
electrostatic potentials to act as short range screened potentials. For the theoretical study of the spectroscopy of a plasma embedded atomic 
system, the screening effects due to the plasma can be conveniently accounted for, in such a scenario, by defining suitable model potentials in 
the atomic Hamiltonian for the corresponding strength of the plasma. The strength of the plasma is defined by a coupling parameter ($\Gamma$) 
that measures the interactions between the particles inside the plasma environment. Debye model \cite{Murillo} is the most conventional approach 
used for studying atomic spectroscopy in low electron-density and high temperature plasma (weakly coupled plasma; i.e $\Gamma$ $<$ 1).

The phenomenon of reduction of ionization potential (IP) of an atom or an ion in the plasma environment is known as ionization potential 
depression (IPD) \cite{Inglis,Feynman,Ecker,Stewart}. Accurate determination of this quantity can infer many useful information such as providing right equation 
of state of plasma, estimating radiative opacity of stellar plasma and inertial confinement fusion plasma, etc. In most of the previous studies, the electronic 
structures of the plasma embedded atomic systems have been investigated using non-relativistic many-body methods. In this work, we consider the 
Dirac-Coulomb (DC) Hamiltonian in the relativistic coupled-cluster (RCC) method, which is explained in the next section, to calculate wave 
functions of the atomic states. Moreover, in most of the previous works the screening effects were taken into account only through the 
nuclear potential. We, however, incorporate screening effects through both the nuclear and electron-electron interactions for more accurate 
description. Its importance was demonstrated recently in a number of works \cite{Ondrej,Jung,Gutierrez,Das2014,Das2016}. In our approach, the two-body 
screening potential is expressed in terms of multiple expansion form and the reduced matrix elements are used for economical 
computation as described in the subsequent section.

In the weakly coupled plasma, the screening effects seen by an electron located at $r_i$ in an atomic system due to the presence of other free 
electrons inside the plasma is accounted by an effective potential given as \cite{Murillo}
\begin{equation}
V_{\mathrm{eff}}(r_i)=  e^{-\mu r_{i}} V_{nuc}(r_i) + \sum_{j \ge i} e ^{-\mu r_{ij}} V_C(r_{ij}),
\label{eqn1}
\end{equation}
where $V_{nuc}(r_i)$ is the usual nuclear potential of electron in the plasma free atomic system and is estimated by consider the Fermi nuclear charge 
distribution and $V_C(r_{ij})=\frac{1}{r_{ij}}$ is the potential due to the two-body Coulomb interactions among the electrons with the Debye screening 
length $1/\mu$. The inverse screening length $\mu$ value is related with the temperature $T$ and electron density $n_e$ of the plasma as
\begin{eqnarray}
\mu= \left [ \frac{4\pi(1+Z)n_e}{k_{\rm{B}} T} \right ]^{1/2}
\label{debye}
\end{eqnarray} 
for the Boltzmann constant $k_{\rm{B}}$ and the nuclear charge $Z$. 

In the multipole expansion, the two-body screened potential can be expressed as
\begin{eqnarray}
V_{ee}(r_i,r_j) &=& \sum_{j \ge i}^{N}\frac{1}{r_{ij}} e^{- \mu r_{ij}} \nonumber \\
       &=& \frac{4\pi}{\sqrt{r_ir_j}}\sum_{k=0}^{\infty}I_{k+\frac{1}{2}}(\mu r_< ) K_{k+\frac{1}{2}} (\mu r_> ) \nonumber \\ 
       && \times \sum_{q=-k}^k Y_{q}^{k\ast}(\theta,\phi)Y^k_{q}(\theta,\phi) ,
\label{veff}
\end{eqnarray}
where $I_{k+\frac{1}{2}}(r)$ and $K_{k+\frac{1}{2}}(r)$ are the modified Bessel functions of the first and second kind, respectively, with $r_> =$ 
max($r_i,r_j$); $r_<=$ min($r_i,r_j$), and $Y^k_{q}(\theta,\phi)$ is the spherical harmonics of rank $k$ with its component $q$. In terms of 
the Racah operator ($C^k_{q}$), the above expression is given by a scalar product as
\begin{eqnarray}
V_{ee}(r_i,r_j) &=& \frac{1}{\sqrt{r_ir_j}}\sum_{k=0}^{\infty} (2k+1) I_{k+\frac{1}{2}} (\mu r_<) K_{k+\frac{1}{2}} (\mu r_> ) \nonumber \\ && \times \mbox{\boldmath$\rm C$}^k(\hat{r}_i) \cdot \mbox{\boldmath$\rm C$}^k(\hat{r}_j).
\end{eqnarray}
In terms of the single particle orbital wave functions ($|\phi(r_i)\rangle$), the above interaction potential in the spherical coordinate 
system can be written as
\begin{eqnarray}
\langle \phi_a \phi_b|V_{ee}(r_i,r_j)| \phi_c \phi_d \rangle &=& (-1)^{j_a -m_a + j_b -m_b} 
\sum_{k,q} (-1)^{k-q} \nonumber \\ & \times & \left (
         \begin{matrix}
         j_a & k & j_c \cr
         -m_a & q & m_c \cr
         \end{matrix}
         \right )  
   \left (
         \begin{matrix}
         j_b & k & j_d \cr
         -m_b & -q & m_d \cr
         \end{matrix}
         \right ) \nonumber \\ & \times & 
         \langle j_a j_b ||{\bf V}_{ee}^k||j_c j_d \rangle,
\end{eqnarray}
where the subscripts $a,b,c$ and $d$ stands for the orbitals, $j$s are the total angular moment and $m$s are their corresponding 
azimuthal components. Here, the allowed $k$ values should be such that $l_a+l_c+k=$even and $l_b+l_d+k=$even for the orbital angular momentum 
quantum number $l$, and they need to satisfy the triangular conditions $|j_a-j_c|\le k \le j_a+j_c$ and $|j_b-j_d|\le k \le j_b+j_d$. 
In the above expression, $\langle j_a j_b || {\bf V}_{ee}^k||j_c j_d \rangle$ is known as the reduced matrix element and is given by
\begin{eqnarray}
\langle j_a j_b||{\bf V}_{ee}^k|| j_c j_d \rangle &=& (-1)^{j_a + j_b + k +1 } 
   \left (
         \begin{matrix}
         j_a & k & j_c \cr
         1/2 & 0 & -1/2 \cr
         \end{matrix}
         \right )  \nonumber \\
  &\times& \left (
         \begin{matrix}
         j_b & k & j_d \cr
         1/2 & 0 & -1/2 \cr
         \end{matrix}
         \right ) \int_0^{\infty} \int_0^{\infty} dr_i dr_j \nonumber \\ 
         & \times & \left [ P_a (r_i) P_c (r_i) +  Q_a (r_i) Q_c(r_i) \right ] \nonumber \\ 
         &\times& (2k+1) \frac{I_{k+\frac{1}{2}} (\mu r_<) K_{k+\frac{1}{2}}(\mu r_>) }{\sqrt{r_ir_j}} \nonumber \\ 
         & \times &  \left [P_b (r_j) P_d (r_j) + Q_b (r_j)Q_d(r_j) \right ],
\end{eqnarray}
where $P(r)$ and $Q(r)$ are the large and small components of the Dirac single particle wave function
\begin{eqnarray}
 |\phi_a(r) \rangle = \frac {1}{r} \left ( \begin{matrix} P_a(r) & \chi^P_{j_a,m_a}(\theta,\phi)  \cr
                                                  iQ_a(r) & \chi^Q_{j_a,m_a}(\theta,\phi) \cr
                                                        \end{matrix}  \right )
\end{eqnarray}
with the respective angular momentum components  $\chi^{P(Q)}_{j_a,m_a}(\theta,\phi)$.
In our formalism, we only use the reduced matrix elements $\langle j_a j_b ||{\bf V}_{ee}^k||j_c j_d \rangle$ to reduce the 
amount of computations.

\begin{table}[t]
\caption{Comparison of ionization potential (IP) from this work with the NIST database \cite{nist}. Absolute differences from the NIST data are given as $\Delta$ in percentage. All these quantities are given in cm$^{-1}$.}\label{tab1}
\begin{center}
\begin{ruledtabular}
\begin{tabular}{l c c c}
Ions & This work & NIST \cite{nist} & $\Delta$ (\%) \\
\hline \\
He I    & 198159.0     & 198310.7  &   0.08 \\
Mg XI   & 14215471.0   & 14209914.7  &  0.04         \\
Fe XXV  & 71276604.2   & 71204137.0  &    0.10       \\
\end{tabular}
\end{ruledtabular}
\end{center}
\end{table}

\begin{table*}[t]
\caption{Comparison of the calculated excitation energies (EEs) in the He-like systems with the NIST database \cite{nist}. Absolute differences from the NIST data are given as $\Delta$ in percentage. All these quantities are given in cm$^{-1}$.}\label{tab2}
\begin{center}
\begin{ruledtabular}
\begin{tabular}{l c c c c c c cccc}
Excited & $J$ & $\pi$ & \multicolumn{2}{c}{He I} &  & \multicolumn{2}{c}{Mg XI}   &  & \multicolumn{2}{c}{Fe XXV} \\
 \cline{4-5} \cline{7-8} \cline{10-11} \\
 state       &    &    & This work & NIST \cite{nist} & & This work & NIST \cite{nist} & & This work & NIST \cite{nist} \\
\hline \\
  & & & \\
$[1s2s] ~ ^1S$ & 1  &  $e$   & 160490.61  & 159855.97  &   & 10748618.12   & 10736136  &  & 53607783.11  &  53527760 \\
$[1s2s] ~ ^3S$ & 0  &  $e$   & 166051.24  & 166277.44  &   & 10846921.10   & 10838778  &  & 53848556.55  &  53781230 \\
$[1s2p] ~ ^3P$ & 2  &  $o$   & 168944.97  & 169086.77  &   & 10843480.69   & 10836388  &  & 53975073.99  &  53896600 \\
               & 1  &  $o$   & 169080.28  & 169086.84  &   & 10840155.63   & 10832818  &  & 53854337.09  &  53777570 \\
               & 0  &  $o$   & 168944.36  & 169087.83  & & 10837603.42     & 10831989  &  & 53823351.52  &  53761280  \\
$[1s2p] ~ ^1P$ & 1  &  $o$   & 170948.69  & 171134.90  &  & 10914104.83  & 10906612  &  & 54121270.71  &  54042490 \\
$[1s3s] ~ ^3S$ & 1  &  $e$   & 183424.43  & 183236.79  &   & 12705049.05  & 12691170  &  & 63546618.39  &  63421700 \\
$[1s3s] ~ ^1S$ & 0  &  $e$   & 184702.01  & 184864.83  &   & 12730224.27  & 12718304  &  & 63612913.04  &  63489000 \\
$[1s3p] ~ ^3P$ & 2  &  $o$   & 185422.75  & 185564.56  &   & 12733512.73   & 12718786  &  & 63612060.10  &  63526300 \\
               & 1  &  $o$   & 185453.86  & 185564.58  &   & 12732443.48   & 12717729  &  & 63584591.83  &  63490700 \\
               & 0  &  $o$   & 185422.57  & 185564.85  & & 12731767.61   & 12717465  &  & 63566942.15  &  63486100 \\
$[1s3d] ~ ^3D$ & 3  &  $e$   & 185960.11  & 186101.55  &   & 12741426.68  & 12733603  &  & 63661108.74  &  63574200 \\
               & 2  &  $e$   & 185960.16  & 186101.55  &   & 12741036.24  & 12733223  &  & 63647766.97  &  63560700 \\
               & 1  &  $e$   & 185966.43  & 186101.59  &   & 12740961.94   & 12733183  &  & 63647055.72  &  63561300 \\  
$[1s3d] ~ ^3D$ & 2  &  $e$   & 185963.12  & 186104.97  &   & 12742065.02   & 12734298  &  & 63662496.91  &  63576500 \\
\end{tabular}
\end{ruledtabular}
\end{center}
\end{table*}

\section{Determination of Atomic Wave Functions}\label{sec3}

 We intend to determine ionization potentials (IPs), excitation energies (EEs) and electron affinities (EAs) of the considered 
He-like ions. The atomic states of the Li-like ions have been constructed in the EA procedure with the He-like ions and their EEs are evaluated 
by subtracting EAs between two states of these ions. For this purpose, we first calculate the wave function of the ground states of the considered He-like ions and 
then pursue with determining IPs, EAs and EEs for these ions and also for the Li-like ions. To carry out these calculations in the RCC theory 
framework, we adopt two distinctly different approaches such as equation-of-motion (EOM) and Fock-space methods. These procedures are described
in the mathematical form below.

The considered DC Hamiltonian in our calculation is given by   
\begin{eqnarray}
H &=& \sum_{i=1}^{N} \left [ c\mbox{\boldmath$\alpha$}_i\cdot \textbf{p}_i+(\beta_i -1)c^2 + e^{-\mu r_{i}} V_{nuc}(r_{i})\right] \nonumber \\
  && +  \frac{1}{2} \sum_{i,j} \frac{e^{-\mu r_{ij}} }{r_{ij}}  
  \label{NSDeq}
\end{eqnarray}
where $\mbox{\boldmath$\alpha$}$ and $\beta$ are the Dirac matrices and $c$ is the velocity of light. 

In the RCC theory, the ground state wave function ($|\Psi_0\rangle$) of a closed-shell atomic system is obtained by expressing
\begin{eqnarray}
|\Psi_0\rangle= e^T |\Phi_0 \rangle,
\end{eqnarray}
where $|\Phi_0 \rangle$ is a mean-field wave function, which is obtained using the Dirac-Fock (DF) method, and $T$ is known as RCC operator
that is responsible for exciting electrons from $|\Phi_0 \rangle$ to excited states when the electron correlation effects, that were neglected
in the DF method, are being included. In the He-like systems, only the singly and doubly excitations are possible which are represented by defining 
$T=T_1 + T_2$. Thus, the above expression naturally gets truncated at
\begin{eqnarray}
|\Psi_0\rangle= (1+T_1+T_2 + \frac{1}{2} T_1^2) |\Phi_0 \rangle .
\end{eqnarray}

The amplitudes of the RCC operators $T$ are obtained by solving the equation
\begin{eqnarray}
\langle \Phi_K | H T_K | \Phi_0 \rangle &=& - \langle \Phi_K | H T_L | \Phi_0 \rangle \nonumber \\
         && - \frac{1}{2} \langle \Phi_K | H T_1^2 | \Phi_0 \rangle,
\label{eqn1}
\end{eqnarray}
where the indices $K$ and $L$ represent for level of excitations with $K \ne L $. The ground state energy ($E_g$) is obtained by
\begin{eqnarray}
\langle \Phi_0| H T_2 + \frac{1}{2} H T_1^2 | \Phi_0 \rangle = E_g .
\label{eqn1}
\end{eqnarray}
 Considering this as our starting point, we can now generate the excited states and their energies, IPs and EAs of He-like and
 Li-like ions in three different steps as described in the following subsections.

\begin{figure*}[t]
\begin{tabular}{ccc}
  \includegraphics[width=6.0cm, height=7.0cm]{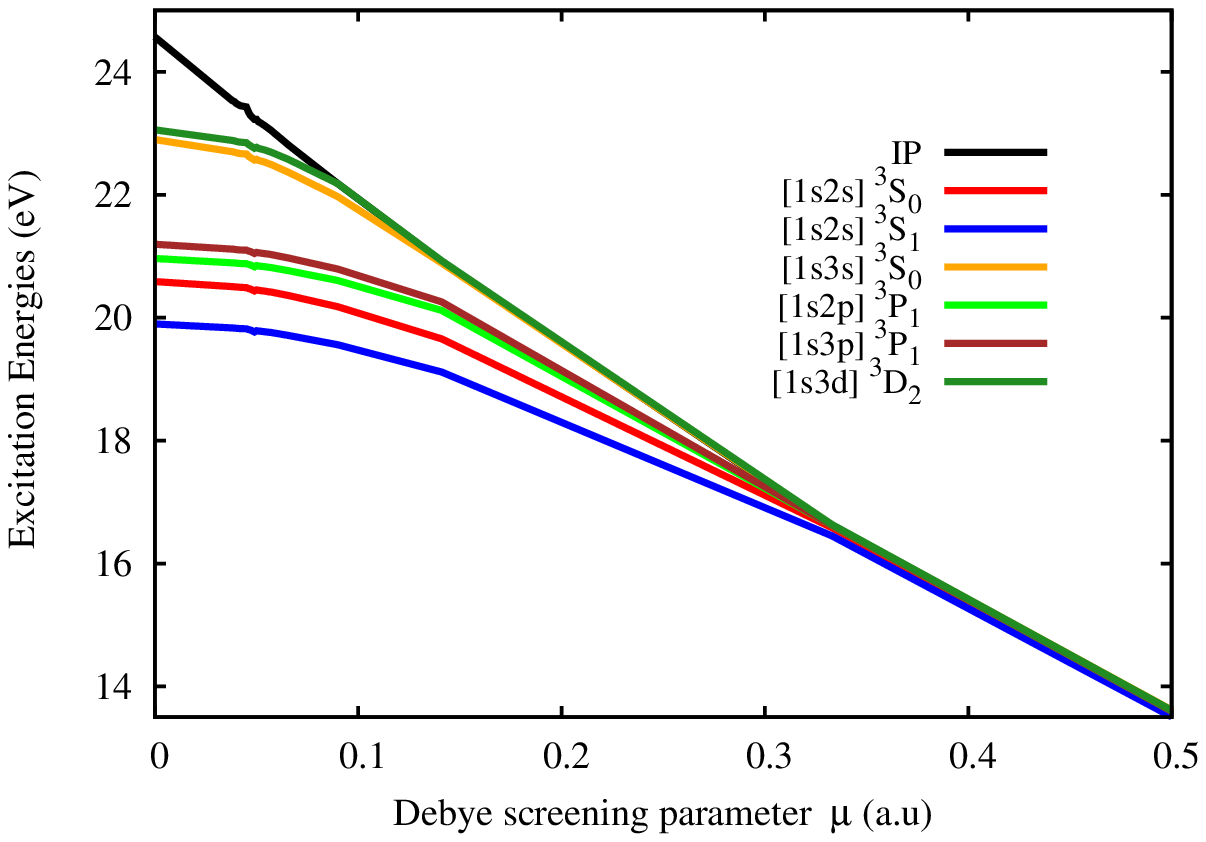}
  & \includegraphics[width=6.0cm, height=7.0cm]{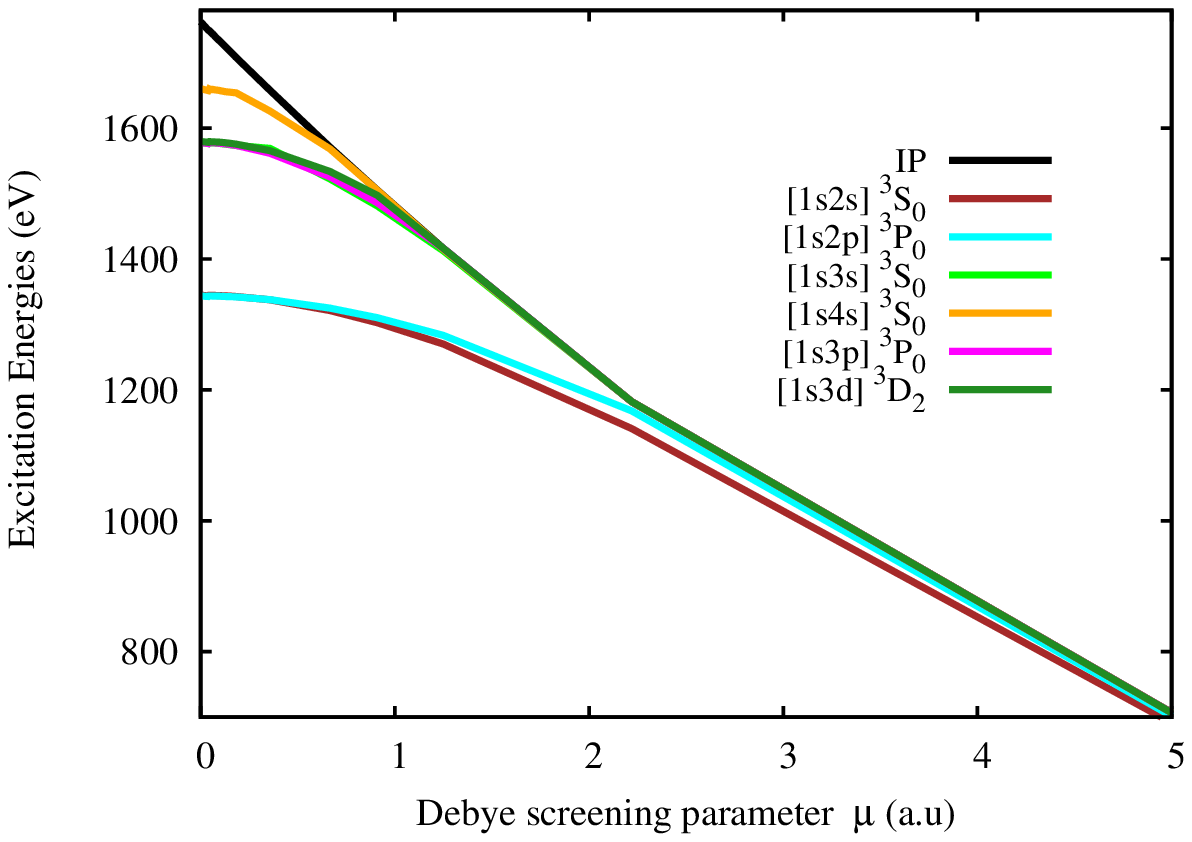} & \includegraphics[width=6.0cm, height=7.0cm]{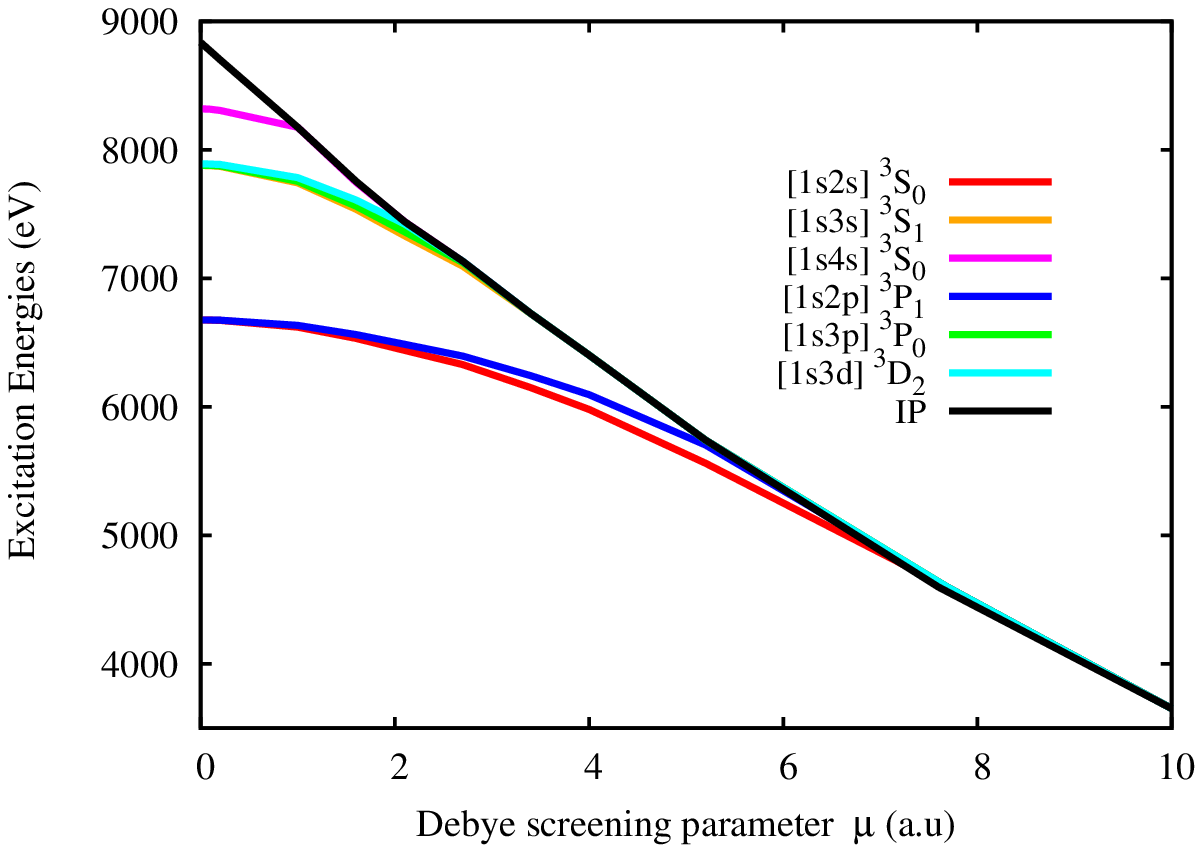} \\
  (a)  & (b) & (c) \\
\end{tabular}
\caption{Variation of EE in (a) He, (b) Mg XI and (c) FeXXV with Debye screening strength $\mu$ in a.u.}
\label{fig1}
\end{figure*}


\subsection{IPs of He-like}

To estimate the IPs of the considered He-like ions, the atomic states of the hydrogen-like (H-like) systems are obtained after removing an 
electron from the $[1s^2]$ configuration using the expression \cite{nandy1}
\begin{eqnarray}
|\Psi_a \rangle = e^T (1+ R_a) |\Phi_a \rangle,
\end{eqnarray}
where the reference state is constructed as $|\Phi_a \rangle = a_a |\Phi_0 \rangle$ with $a_a$ is the corresponding annihilation operator 
and $R_a$ is another RCC operator that takes care of the extra correlation effects accounted through the removed $1s$ orbital electron while 
generating $|\Psi_0\rangle$. The $R_a$ operator can also give rise only the singles and doubles excitations and denoted by $R_a=R_{1a}+ R_{2a}$. The IP
($E_a$) and amplitude solving equations for the $R_a$ wave operators are given by
\begin{eqnarray}
\langle \Phi_a| (H e^T) \{1+R_a\}|\Phi_a\rangle &=& E_a \\
\langle \Phi^b_a| [(H e^T)- E_a] R_a |\Phi_a \rangle &=&  - \langle \Phi^b_a| H e^T |\Phi_a \rangle \label{eq26}
\end{eqnarray}
and 
\begin{eqnarray}
\langle \Phi^{pb}_{da}| [(H_N e^T)-E_a] R_a |\Phi_a \rangle &=&  -\langle \Phi^{pb}_{da}| H e^T |\Phi_a \rangle, \ \ \ \label{eq27}
\end{eqnarray}
where $|\Phi_a^b \rangle$s are the singly excited configurations from $|\Phi_a \rangle$ constructed replacing an occupied orbital $a$ 
by another occupied orbital $b$ and $|\Phi^{pb}_{da} \rangle$s denote the doubly excited configurations from $|\Phi_a \rangle$, constructed 
replacing an occupied orbital $a$ by orbital $b$ and exciting an electron from the occupied orbital $d$ to virtual orbital $p$. The above
non-linear equations are solved self-consistently along with the energy evaluating equation. 

\subsection{EEs of He-like}
 
 The excited states ($|\Psi_K(J, \pi) \rangle$) with angular momentum $J$ and parity $\pi$ is obtained by operating excitation operators 
 $\Omega_K$ on $|\Psi_0 \rangle$ as 
\begin{eqnarray}
 |\Psi_K(J, \pi) \rangle = \Omega_K(J, \pi) |\Psi_0\rangle ,
\end{eqnarray}
where $K$ corresponds to level of excitations (in the present case it is naturally truncated at the double excitations). Thus, the eigenvalue ($E_L$) and 
eigenfunctions for the $L^{th}$ excited state are obtained by diagonalizing the equation
\begin{eqnarray}
\langle \Phi_L (J, \pi) | H^{eff}  \Omega_K (J, \pi) |\Phi_0 \rangle = E_L \langle \Phi_L (J, \pi) | \Omega_L (J, \pi) |\Phi_0 \rangle , \nonumber
\end{eqnarray}
where $H^{eff}=(H+HT_1+HT_2 +\frac{1}{2}HT_1^2)$. It is obvious from the above equation that it is imperative to project $|\Phi_L\rangle$ for 
a definite value of $J$ and $\pi$ in order to get solutions for the respective states and the equation needs to be solved self-consistently 
for both the singles and doubles excitations. Using the Davidson's diagonalization algorithm, only the solutions for the lower energy levels are 
obtained for our interest.

 The above method is routinely used in the quantum chemistry, however it is developed by us recently for atomic systems with
spherical coordinate system \cite{nandy2}. The bottle-neck for using this method in the spherical coordinate description is that
both the Hamiltonian $H$ and the $T$ are expressed in terms of multiple expansion form resulting following type of tensor 
products 
\begin{eqnarray}
\langle J \pi || [{\bf t}^{k_1} {\bf u}^{k_2}]^K || J' \pi  \rangle = (2K+1)^{1/2} (-1)^{J+J'+K} \sum_{J''}   \nonumber \\
                               \times   \left \{ 
                                           \begin{matrix}
                                           k_1 & k_2 & K \cr
                                           J' & J & J'' \cr
                                           \end{matrix}
                                         \right \}
                                         \langle J \pi ||{\bf t}^{k_1} ||J''  \pi \rangle \langle J''  \pi ||{\bf u}^{k_2} ||J'  \pi \rangle .
\end{eqnarray}
Therefore, the resultant operator becomes another tensor with different rank. This complicates to account for the angular momentum coupling 
between the operators with different allowed intermediate $J''$ states and storing them optimally for carrying out computations. 
However, this approach refrains from performing calculations using $m$ sublevels. It, thus, allows to embody a large configuration 
space for more accurate calculations.

\begin{table}
\caption{Comparison of EAs of the electrons in the low-lying excited bound states of Mg XI and Fe XXV ions with the NIST database 
\cite{nist}. Absolute differences from the NIST data are given as $\Delta$ in percentage. All these quantities are given in cm$^{-1}$.}\label{tab3}
\begin{center}
\begin{ruledtabular}
\begin{tabular}{l c c c}
State & \multicolumn{2}{c}{This work}  &  NIST \cite{nist} \\
 \cline{2-3} \\
      &   EA   &  EE   &  EE \\
\hline \\
\multicolumn{4}{c}{Mg XI} \\
$2S$    & 2962134.64   & 0   &  0  \\
$2P_{1/2}$ & 2802991.11  & 159143.53  & 160015  \\
$2P_{3/2}$ & 2798659.31 &  163475.33  & 163990  \\
$3S$  & 1278664.89 &   1683469.75   &   1682700  \\
$3P_{1/2}$ & 1231022.75 & 1731111.89  & 1726520   \\
$3P_{3/2}$  & 1229758.76 & 1732375.88 & 1727830  \\
$3D_{3/2}$   & 1217780.52 & 1744354.12  & 1743500  \\
$3D_{5/2}$  & 1217617.80 &  1744516.84  & 1743890  \\
     & & & \\
    \multicolumn{4}{c}{Fe XXV} \\
$2S$    &  16500309.62  & 0   & 0  \\
$2P_{1/2}$   & 16110730.72 &  389578.90  & 391983 \\
$2P_{3/2}$  & 15977474.34 & 522835.28 & 520757  \\
$3S$  & 7177704.50 &  9322605.12  &  9272500  \\
$3P_{1/2}$   & 7109738.12  &  9390571.50  &  9378200  \\
$3P_{3/2}$  & 7069659.55 & 9430650.07 & 9417100  \\
$3D_{3/2}$   & 7032196.18  &  9468113.44 & 9459000  \\
$3D_{5/2}$  & 7020171.90 &  9480137.72 & 9472600  \\   
\end{tabular}
\end{ruledtabular}
\end{center}
\end{table}

\begin{figure}[t]
  \includegraphics[width=9.0cm, height=7.0cm]{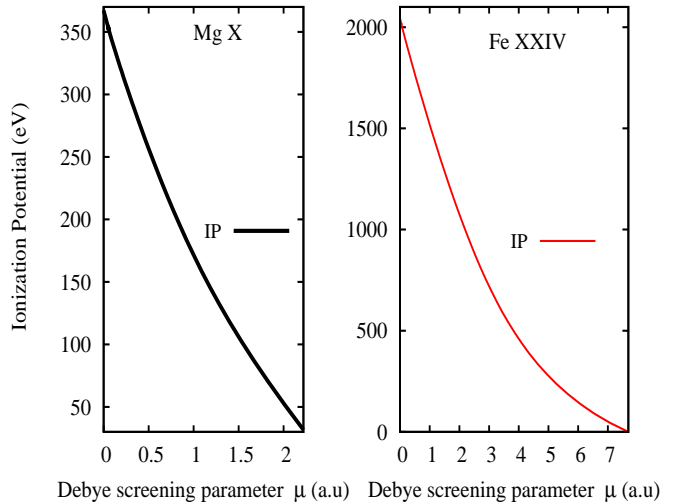}
\caption{Variation of IPs of electrons in various bound states of the Fe XXV and Mg X ions with the Debye screening strength ($\mu$) in a.u.}
\label{fig2}
\end{figure}

\begin{figure}[t]
  \includegraphics[width=9.0cm, height=7.0cm]{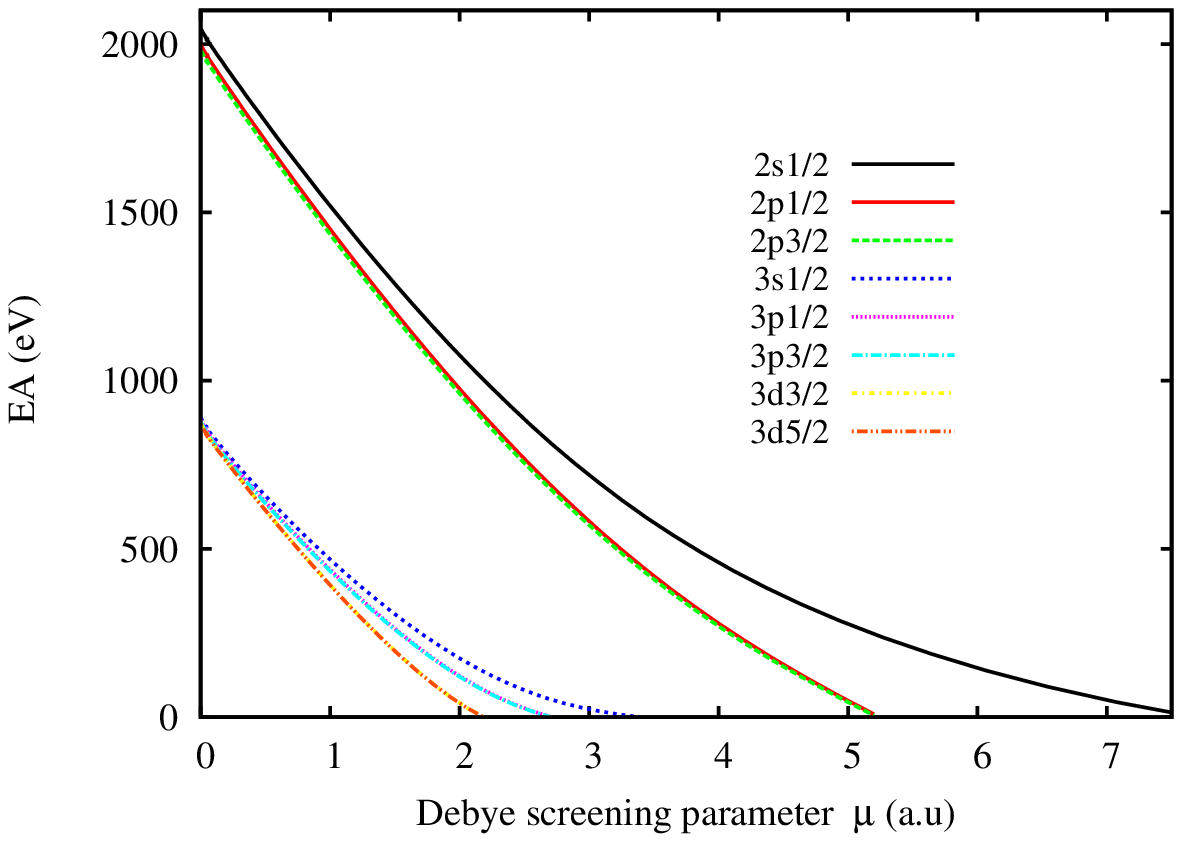}
\caption{Variation of EAs of electrons in various bound states of the Mg XI ion with the Debye screening strength ($\mu$) in a.u.}
\label{fig3}
\end{figure}

\begin{figure}[t]
  \includegraphics[width=9.0cm, height=7.0cm]{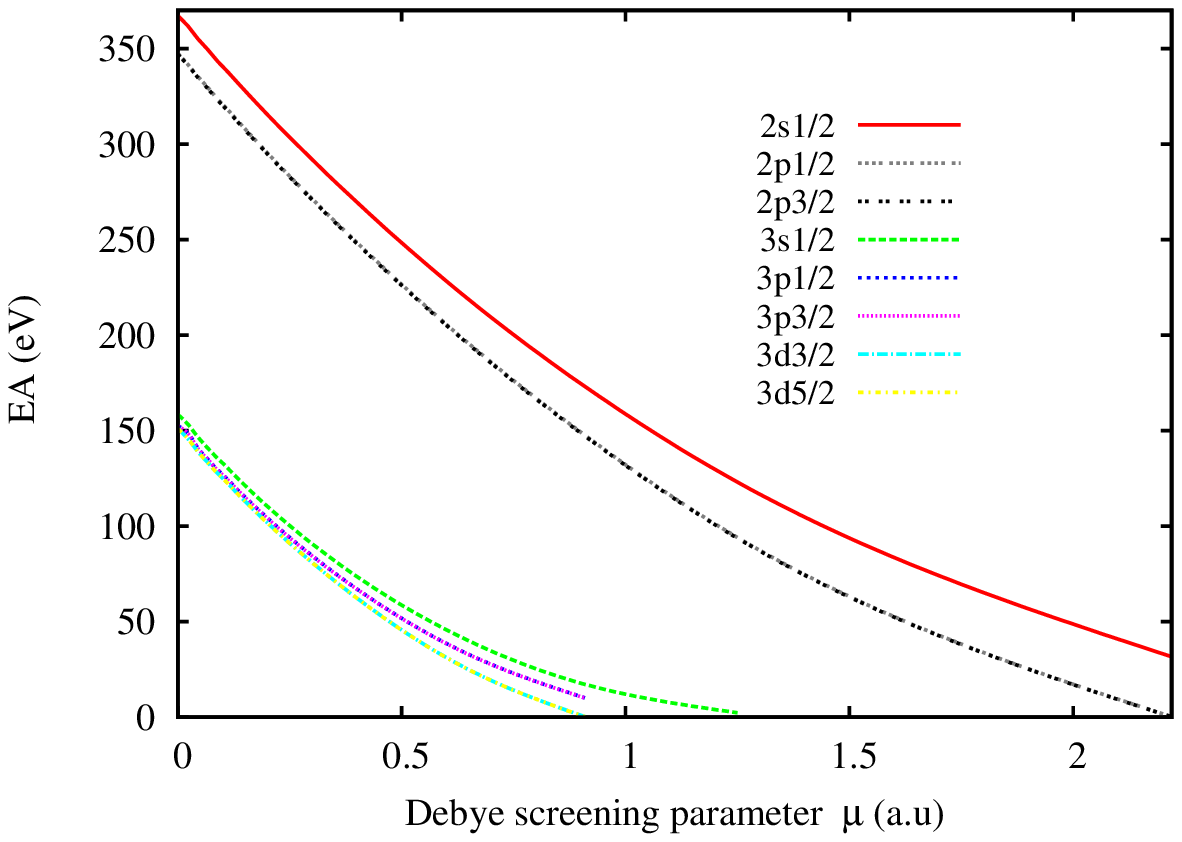}
\caption{Variation of EAs of electrons in various bound states of the Fe XXIV ion with the Debye screening strength ($\mu$) in a.u.}
\label{fig4}
\end{figure}

\begin{table*}[t]
\caption{Variation of electron EAs in cm$^{-1}$ of various low-lying excited bound states in the Mg XI ion 
with $\mu$ values.}\label{tab4}
\begin{center}
\begin{ruledtabular}
\begin{tabular}{l c c c c c c c c}
$\mu$ & 2S$_{1/2}$ &  2P$_{1/2}$  &  2P$_{3/2}$ & 3S$_{1/2}$ & 3P$_{1/2}$ & 3P$_{3/2}$ & 3D$_{3/2}$ & 3D$_{5/2}$ \\
\hline
    0  & 2962134.64 & 2802991.11 & 2798659.31 & 1278664.89 & 1231022.75 & 1229758.75 & 1217984.45 & 1217617.79\\       
  0.039   & 2877517.44 & 2717070.18 & 2713118.82 & 1195955.88 & 1151893.96 & 1150730.12 & 1134767.13 & 1134408.81\\    
  0.044   & 2865436.72 & 2706144.90 & 2701815.34 & 1183545.66 & 1135822.49 & 1134561.61 & 1122350.07 & 1121984.42 \\  
  0.046   & 2862227.63 & 2702921.55 & 2698592.14 & 1180442.37 & 1132710.10 & 1131449.44 & 1119208.60 & 1118843.01\\        
  0.047   & 2859049.70 & 2699719.89 & 2695390.59 & 1177366.80 & 1129622.12 & 1128361.70 & 1116090.56 & 1115725.26 \\  
  0.048   & 2857107.11 & 2697789.18 & 2693460.00 & 1175496.83 & 1127761.92 & 1126501.62 & 1114211.94 & 1113846.69 \\  
  0.052   & 2849628.38 & 2690279.80 & 2685951.01 & 1168289.37 & 1120538.32 & 1119278.56 & 1106913.61 & 1106548.50\\   
  0.053   & 2847926.37 & 2688575.81 & 2684247.13 & 1166654.10 & 1118901.95 & 1117642.32 & 1105259.52 & 1104894.54 \\
  0.054   & 2844338.05 & 2684945.50 & 2680617.04 & 1163194.64 & 1115418.66 & 1114159.29 & 1101737.77 & 1101372.82\\
  0.056   & 2840269.78 & 2680873.20 & 2676544.99 & 1159294.53 & 1111517.17 & 1110258.09 & 1097791.71 & 1097426.88\\ 
  0.057   & 2838128.56 & 2678729.90 & 2674401.77 & 1157245.57 & 1109465.99 & 1108207.09 & 1095716.53 & 1095351.81\\
  0.058   & 2834752.03 & 2675333.08 & 2671005.16 &1154007.94  & 1106218.36 & 1104959.73 & 1092430.01 & 1092065.35\\
  0.061   & 2828778.64 & 2669347.80 & 2665020.32 &1148302.26  & 1100505.72 & 1099247.64 & 1086646.35 & 1086281.85\\
  0.063   & 2824348.10 & 2664891.18 & 2660563.96 &1144070.06  & 1096260.07 & 1095002.35 & 1082345.85 & 1081981.47\\
  0.065   & 2821954.33 & 2662504.39 & 2658177.30 & 1141796.29 &  1093989.19& 1092731.71 & 1080044.89 & 1079680.59\\       
  0.076    & 2796870.11 & 2637257.84 & 2632932.55 &1117985.71  & 1070087.05 & 1068832.04 & 1055794.16 & 1055430.77\\           
  0.095    & 2758527.64 & 2598693.02 & 2594371.15 &1082006.77  & 1033997.97 & 1032747.58 & 1019068.14 & 1018706.34\\      
  0.125     & 2697116.20 & 2536745.01 & 2532429.44 & 1025375.54 & 977086.88 & 975845.75  & 960876.71 & 960518.02 \\   
  0.181    & 2582557.37 & 2420906.92 & 2416609.72 &923068.56   &  874157.32 & 872392.49 & 854747.08 & 854396.59 \\
  0.357     & 2250918.79 & 2082777.80 & 2078575.58 &651203.19   & 599695.24  & 598587.08  & 565799.09  & 565487.95\\
  0.666     & 1738945.01 & 1552044.13 & 1548127.51 &308087.58   & 251006.98  & 250189.58  & 185554.21  & 185350.79\\
  0.909    & 1396629.83 & 1190984.40 & 1187379.29 &138853.91   & 81660.82   &  81140.58  & 1924.47    &  1837.01 \\      
  1.25    & 992478.50  & 758867.13  & 755805.58  & 18620.32   &            &            &            &      \\  
  2.2    &  255447.24 &   1861.67  &  1110.17   &            &            &            &             &       \\
\end{tabular}
\end{ruledtabular}
\end{center}
\end{table*}

\begin{table*}[t]
\caption{Variation of electron EAs (in cm$^{-1}$) of various low-lying excited bound states in the Fe XXV ion with 
$\mu$ values.}\label{tab5}
\begin{center}
\begin{ruledtabular}
\begin{tabular}{l c c c c c c c c c c c c c c c c}
$\mu$ & &2S$_{1/2}$ & & 2P$_{1/2}$ & &  2P$_{3/2}$& & 3S$_{1/2}$& & 3P$_{1/2}$& & 3P$_{3/2}$ & &3D$_{3/2}$ & &3D$_{5/2}$ \\
\hline  
  0        & & 16500309.61  && 16110730.71 & & 15977474.33 & & 7177704.50 & & 7109738.12 & & 7069659.54 & & 7032196.17 & & 7020171.90\\
  0.040    & & 16285972.28 & & 15896184.71 & & 15762938.40 & & 6964747.13 & & 6896541.36 & & 6856476.85 & & 6818678.65 & & 6806659.25\\        
  0.041   & & 16282838.78  && 15893081.75 & & 15759835.68 & & 6961663.88 & & 6893477.82 & & 6853413.69 & & 6815605.82 & & 6803586.56\\           
  0.042    & & 16275962.91  && 15886216.76 &  &15752971.38 & & 6954895.37 & & 6886701.75 & & 6846638.57 & & 6808808.71 & & 6796789.78\\       
  0.044    & & 16268176.81 & & 15878393.56 & & 15745148.95 & & 6947206.18 & & 6878983.04 & & 6838920.93 & & 6801065.22 & & 6789046.64\\       
  0.045    & & 16263991.13 & & 15874220.57 & & 15740976.42 & & 6943081.82 & & 6874867.45 & & 6834805.94 & & 6796936.05 & & 6784917.70\\    
  0.048    & & 16246362.47 & & 15856529.28 & & 15723287.05 & & 6925700.64 & & 6857430.08 & & 6817371.20 & & 6779438.50 & & 6767421.13\\      
  0.049    & & 16242746.40 & & 15852921.16 & & 15719679.28 & & 6922146.07 & & 6853876.02 & & 6813817.72 & & 6775871.68 & & 6763854.42\\    
  0.050    &  &16234837.54 & & 15844942.18 & & 15711746.31 & & 6914363.11 & & 6846063.50 & & 6806006.47 & & 6768030.38 & & 6756013.61\\    
  0.052    &  &16225549.25 & & 15835690.29 & & 15702450.45 & & 6905228.78 & & 6836913.67 & & 6796858.14 & & 6758845.75 & & 6746829.45\\  
  0.053    & & 16220908.52 & & 15831016.67 & & 15697777.39 & & 6900653.27 & & 6832315.79 & & 6792261.04 & & 6754229.90 & & 6742213.94\\ 
  0.054    & & 16212954.54 & & 15823073.30 & & 15689834.99 & & 6892841.22 & & 6824504.33 & & 6784450.96 & & 6746387.30 & & 6734371.80\\      
  0.055    & & 16208838.17 & & 15818930.24 & & 15685692.41 & & 6888798.29 & & 6820431.34 & & 6780378.66 & & 6742297.67 & & 6730282.47\\       
  0.057    & & 16199505.94 & & 15809587.98 & & 15676351.45 & & 6879635.91 & & 6811251.08 & & 6771200.11 & & 6733079.13 & & 6721064.49\\      
  0.059    & & 16189189.07 & & 15799259.90 &  &15666024.72 & & 6869510.76 & & 6801107.91 & & 6761058.85 & & 6722892.28 & & 6710878.29\\      
  0.060    & & 16183966.35 & & 15794016.17 & & 15660781.74 & & 6864387.66 & & 6795960.40 & & 6755912.33 & & 6717722.03 & & 6705708.35\\      
  0.066     & & 16151950.92 & & 15761928.26 & & 15628698.47 & & 6833007.28 & & 6764496.44 & & 6724454.82 & & 6686110.72 & & 6674099.37\\      
  0.1     & & 15979834.43 & & 15589280.75 & & 15456083.86 & & 6665398.68 & & 6596230.00 & & 6556233.54 & & 6516813.27 & & 6504817.77\\    
  0.2      & & 15471586.71 & & 15078244.96 & & 14945223.17 & & 6180753.69 & & 6108227.26 & & 6068465.67 & & 6023455.17 & & 6011543.01\\   
  1.0      &  &11811949.59 & & 11340008.80 & & 11212053.52 & & 3145905.15 & & 2996513.88 & & 2962885.61 & & 2780061.10 & & 2770344.42\\  
  1.6     & & 9255002.30  & & 8659403.99  &  &8539838.58  & & 1531754.65 & & 1304912.74 & & 1280461.25 & & 925126.45  & & 918731.58\\            
  2.18    & & 7544279.71  & & 6829743.19  & & 6718801.51  & & 722812.95  & & 466509.67  & & 450525.30  & & 13083.88   &  &10165.00\\     
  2.7     & & 6041681.55  & & 5197119.63  & & 5096434.10  & & 233105.76  & & 12504.03   & & 7107.10    &  &           &  &        \\
  0.293    & & 4307604.21  & & 3285607.86  & & 3201171.92  & & 331.21     & &            & &            &  &           &  &        \\
  3.41     & & 3130516.75  & & 1980734.90  &  &1911462.43  & &            & &            & &            &  &           &  &       \\
  5.26     & & 1296811.90  &  &70094.38    & & 42621.76    & &            & &            & &            &  &           &  &        \\ 
  7.69     &  &790.73      &  &            &  &            & &            & &            & &            &  &           &   &       \\         
\end{tabular} 
\end{ruledtabular}
\end{center}
\end{table*}

\subsection{EAs of He-like and EEs of Li-like }

The EAs of the considered He-like ions are obtained by appending electrons in the valence $s$, $p$, $d$ orbitals of the $[1s^2]$ 
configuration. This can also give atomic states of the Li-like ions. In the Fock-space formalism of RCC theory the corresponding states are
expressed as \cite{Das2014,Das2016,bijaya}
\begin{eqnarray}
|\Psi_v \rangle = e^T (1+ S_v) |\Phi_v \rangle,
\end{eqnarray}
where the reference state is constructed as $|\Phi_v \rangle = a_v^{\dagger} |\Phi_0 \rangle$ with $a_v$ representing creation of the valence
orbital and $S_v$ is the RCC operator that takes into account the correlation effects seen by the valence electron interacting with the other 
occupied orbitals. In this case too, the $S_v$ operator can account only the singles and doubles excitations which is denoted by $S_v=S_{1v}+ 
S_{2v}$. The EA ($E_v$) and amplitude solving equations for the $S_v$ wave operators are given by
\begin{eqnarray}
\langle \Phi_v| (H e^T) \{1+S_v\}|\Phi_v\rangle &=& E_v \\
\langle \Phi^p_v| [(H e^T)- E_v] S_v |\Phi_v \rangle &=&  - \langle \Phi^p_v| H e^T |\Phi_v \rangle 
\end{eqnarray}
and 
\begin{eqnarray}
\langle \Phi^{pq}_{vb}| [(H_N e^T)-E_v] S_v |\Phi_v \rangle &=&  -\langle \Phi^{pq}_{vb}| H e^T |\Phi_v \rangle,
\end{eqnarray}
where $|\Phi_v^p \rangle$ are the singly excited configurations from $|\Phi_v \rangle$ constructed replacing the valence orbital $v$ 
by an virtual orbital $p$ and $|\Phi^{pq}_{vb} \rangle$ denotes the doubly excited configurations from $|\Phi_v \rangle$, constructed 
replacing simultaneously the valence orbital $v$ by a virtual orbital $p$ and exciting an electron from the occupied orbital $b$ to 
virtual orbital $q$. These non-linear equations are also solved self-consistently along with the energy evaluating equation. By taking 
differences between EAs of different orbitals, EEs of higher excited states of Li-like ions are determined.

\section{Results and Discussion}\label{sec4}
We have adopted RCC method to compute IPs and EEs of He and He-like Mg and Fe ions in the weak plasma environment using Debye plasma model. EAs of Li-like Mg and Fe ions are also determined by extending calculations of their He-like ions. In order to validate our calculations, we
have also performed calculations of the above quantities in the plasma free environment considering $\mu=0.0$ and compared them against their 
corresponding values quoted in NIST database \cite{nist}. Most of NIST data are obtained from high precision calculations using more accurate numerical methods that take into account the contributions from quantum electrodynamics (QED) rigorously. IPs of plasma free He I, Mg XI and Fe XXV systems are presented in Table \ref{tab1} along with the NIST data. Our calculations also agree well with the NIST data and they are found to be sub-one percent accurate. Discrepancies in the He-I system are mainly due to poor description of nuclear charge distribution, while they are mainly due to the neglected QED corrections in the other ions. Similarly, EEs of many low-lying transitions of these systems are given in Table \ref{tab2}. Nevertheless, we intend to demonstrate the trend of IPs and EEs of these ions with different plasma strengths ($\mu$). In Fig. \ref{fig1}, we show the trends of IPs and EEs of many representative states with different $J$ values and parities. As seen in the figures, the trends of these 
quantities in different states differ from He I to highly charged ions. It is noticed that in the plasma environment the energy level structures are different in these isoelectronic systems and the plasma screening effects in these states also behave differently. IPs decrease gradually till they become zero for some critical value of $\mu$ (say $\mu_c$) beyond which the states transform to continuum. As mentioned before, the corresponding IP beyond which instability occurs is known as IPD. Variations in IPs with $\mu$ values are shown in black line (Fig \ref{fig1}) at certain plasma, while EEs of the higher excited states are shown in color lines  (Fig \ref{fig1}). As the fine structure splitting between the degenerate levels are unaffected by the increasing plasma strength, we have plotted some selected states among the excited degenerate levels. From the figure, it is evident that the EEs of different levels gradually decrease with increasing plasma screening and finally merge into the continuum at 
particular $\mu_c$. This variation is more rapid near ionization limit. For example, as seen in Fig \ref{fig1}(a), the critical $\mu_c$ value for the $[1s3d] ^3D$ state of He I is approximately 0.1 a.u. Similar trends are also obtained in Mg XI, as shown in Fig \ref{fig1}(b), and in Fe XXV, as shown in Fig. \ref{fig1}(c). In these ions, we observe that $\mu_c$ values of these ions for different configurations are in the order $[1s4s] < [1s3d] < [1s3p] < [1s2p] < [1s2s]$. As a result, the number of bound states of the embedded plasma ions get reduced in comparison to the plasma free systems.

After investigating IPs and EEs of the He-like systems, we now present EAs and EEs of the Li-like ions. Again, we have also 
calculated these quantities for the plasma free systems considering $\mu=0$ in order to compare them against the previously 
reported values 
quoted in the NIST database \cite{nist}. They are given in Table \ref{tab3} which shows that our calculations are in good agreement 
with the NIST data. This assures that our calculations with plasma screening effects will also be of similar accuracy within the Debye 
model framework. In Fig. \ref{fig2}, we have shown the variation of IPs of Mg X and Fe XXIV with $\mu$. The figure clearly shows 
that the IPD of these plasma embedded ions are similar to that of He-like Mg and Fe ions. In Figs. \ref{fig3} and \ref{fig4}, we have 
plotted the variation of EAs of Mg X and Fe XXIV with 
increasing plasma strength $\mu$. This clearly demonstrates that the EAs of both of the plasma embedded ions decrease with increasing 
$\mu$ values, but its rate is slower than change in IPs of the respective He-like ions. For quantitative estimate of these quantities,
we have also given EAs of the low-lying 2S$_{1/2}$, 2P$_{1/2,3/2}$, 3S$_{1/2}$, 3P$_{1/2,3/2}$, and 3D$_{3/2,5/2}$ states
of the Mg X and Fe XXIV ions in Tables \ref{tab4} and \ref{tab5}, respectively, for selective values of $\mu$. This information 
will be quite useful for the astrophysical plasma and tokamak plasma for diagnostic of their processes. From the 
tables, we can infer that as we increase the strength of plasma, the bound atomic orbitals  migrate towards the continuum 
making the ions unstable in the plasma. From Fig. \ref{fig3}, we can also infer that the atomic states with same principal
quantum number $n$ and different orbital quantum number $l$, i.e. the [2S,2P$_{1/2,3/2}$], [3S,3P$_{1/2,3/2}$], and 
[3D$_{3/2,5/2}]$ states in Mg X, are almost degenerate at lower screening strengths but they gradually split farther with the
increasing strength of the plasma. On the otherhand the fine structure splitting between the 2P$_{1/2,3/2}$, 3P$_{1/2,3/2}$ and 
3D$_{3/2,5/2}$ do not affect much with the increasing value of $\mu$. Similar trends are also seen in the Fe XXIV ion as shown in
Fig. \ref{fig4}.   

\section{Concluding remarks}\label{sec6}

We have applied equation-of-motion and Fock-space coupled-cluster methods in the relativistic framework to investigate the
trends of ionization potential and excitation energies of He-like and Li-like Mg and Fe ions in Debye plasma environment. 
We have considered Debye screening both in the nuclear and two-body Coulomb interaction potential and performed the 
calculations by carrying out multipole expansion approach in the spherical coordinate system. We found that the ionization potentials 
in the He-like systems vary faster than electron affinities of the Li-like ions. We have also given explicitly electron 
affinities of the considered Li-like ions for some intermediate values of plasma strength which can be used for 
diagnostic of plasma processes. We also observe that atomic energy levels have smaller energy gap for higher plasma strength 
while their differences increase among the states having same principal and different orbital quantum numbers. However, 
fine structure splitting among different states are least affected with increasing strength of plasma. These results will be 
useful in interpreting the laboratory and astrophysical plasma.  

\section*{Acknowledgment}
M. Das acknowledge Department of Science and Technology, Government of India for financial support vide reference No.SR/WOS-A/PM-10/2016 (G) under Women Scientist Scheme to carry out this work. Computations were carried out using Vikram-100 HPC cluster at Physical Research Laboratory, Ahmedabad.

\end{document}